\documentclass[aps,preprint,superscriptaddress,showpacs]{revtex4-1}

\usepackage{times}
\usepackage{amsmath,bm,amsfonts}
\usepackage{dcolumn}% Align table columns on decimal point
\usepackage{graphicx}
\usepackage{latexsym}

\usepackage{color}

\begin{document}
%\preprint

\title{New Candidates for Topological Insulators : Pb-based chalcogenide series}

\author{Hosub \surname{Jin}}

\author{Jung-Hwan \surname{Song}}
\email{jhsong@pluto.phys.northwestern.edu}

\author{Arthur J. \surname{Freeman}}

\affiliation{Department of Physics and Astronomy, Northwestern University,
    Evanston, Illinois 60208, USA}

\author{Mercouri G. \surname{Kanatzidis}}

\affiliation{Department of Chemistry, Northwestern University,
    Evanston, Illinois 60208, USA}

%\date{\today }

\begin{abstract}
To date, Bi$_2$Se$_3$ is known as the best three-dimensional high temperature
topological insulator with a large bulk band gap.
%one simple surface state Dirac cone, and a simple surface geometry from the layered crystal
%structure and van der Waals type interlayer interactions.
Here, we theoretically predict that the series of Pb-based layered chalcogenides,
Pb$_n$Bi$_2$Se$_{n+3}$ and Pb$_n$Sb$_2$Te$_{n+3}$, are possible new candidates for topological insulators.
As $n$ increases, the phase transition from a topological insulator to a band insulator is found to occur
between $n=2$ and 3 for both series. Significantly, among the new topological insulators, we found a bulk
band gap of 0.40eV in PbBi$_2$Se$_4$ which is one of the largest gap topological insulators, and that
Pb$_2$Sb$_2$Te$_5$ is located in the immediate vicinity of the topological phase boundary, making
its topological phase easily tunable by changing external parameters such as lattice constants.
Due to the three-dimensional Dirac cone at the phase boundary, massless Dirac fermions also may be
easily accessible in Pb$_2$Sb$_2$Te$_5$.
\end{abstract}

\pacs{ }

\maketitle

Topological insulators, distinguished from normal band insulators by a nontrivial Z$_2$ topological
number and topologically protected surface states, have attracted great attention due to their
significance both for applications and for fundamental research on a new quantum state of matter
~\cite{2010hasan,qi:33}. Since the suggestion of a quantum spin Hall effect in a honeycomb lattice
~\cite{PhysRevLett.95.226801,PhysRevLett.95.146802}, which is a time-reversal pair of Haldane models
~\cite{PhysRevLett.61.2015} and induced by spin-orbit interactions, several topological insulator
materials were predicted and observed~\cite{Bernevig12152006,MarkusKonig11022007,PhysRevLett.98.106803,
Hsieh2008,Roushan2009,Xia2009,Zhang2009}. Efforts to find new topological insulators are still being
made~\cite{Chadov2010,Lin2010}. A large spin-orbit coupling (SOC) strength is essential to realizing
non-trivial topological band structures.

To classify the Z$_2$ topological number, we adopted the method proposed in ref.~\cite{PhysRevB.78.165313},
which studied the topological phase transition by making a variation of external parameters such as the
SOC strength or lattice constants. In our work, we varied the SOC strength ($\lambda_{\mathrm{SO}}$) from 0
to the real values of the systems ($\lambda_{0}$), and investigated whether the topological phase transition
occurs. In three-dimensional (3D) topological insulator materials, a band crossing between conduction and
valence bands should occur during this process: the 3D Dirac cone appears at the transition point between the
band insulator (BI) and topological insulator (TI).

There are several advantages of this method. First, it is intuitive and only needs a direct observation
of the Dirac cone at the transition point. Second, it does not require a heavy computational cost because of
the simple unit-cell calculations. The surface calculations to see the topologically protected surface state
are also one of alternative ways to determine the topological phase. The degrees of complexity
of such calculations are, however, grater than the method we employed here. Third, this method can be
applied to every system regardless of the existence of inversion symmetry. Lastly, the critical SOC value
at which the band crossing occurs shows how far the system is located from the phase boundary of BI and TI.

By using this method, we found that the series of Pb-based chalcogenides, Pb$_n$Bi$_2$Se$_{n+3}$ and
Pb$_n$Sb$_2$Te$_{n+3}$, are possible new candidates for TI materials and the topological phases are
changed from TI to BI with increasing $n$. Among the above series, we focus on the material near the
phase boundary between BI and TI due to the possibility of tuning the topological phase by changing
external parameters.
%Especially, the material near the phase boundary is focused on due to
%the possibility of tuning the topological phase by changing external parameters.

To investigate the electronic structures and topological phases, first-principles calculations
were performed using the full-potential linearized augmented plane wave method\cite{PhysRevB.24.864}
with the gradient-corrected Perdew, Burke, and Ernzerhof form of the exchange-correlation functional
~\cite{PhysRevLett.77.3865}. The core states and the valence states were treated fully relativistically
and scalar relativistically, respectively. The calculations were carried out with the experimental lattice
parameters for bulk PbBi$_2$Se$_4$, PbSb$_2$Te$_4$ and Pb$_2$Bi$_2$Se$_5$, and with the fully optimized
geometry for the others, because crystal structures of $n=1$ compounds and Pb$_2$Bi$_2$Se$_5$ were reported
experimentally~\cite{agaev68,shelimova04,agaev66} and those of other $n$'s are designed here theoretically.
%The optimized and experimental lattice constants in PbBi$_2$Se$_4$, Pb$_2$Bi$_2$Se$_5$,
%and PbSb$_2$Te$_4$ show a good agreement, where the difference between them is as usual within 1$\sim$3$\%$.

%\emph{\textbf{PbBi$_2$Se$_4$ and PbSb$_2$Te$_4$}}

PbBi$_2$Se$_4$ and PbSb$_2$Te$_4$ have the rhombohedral crystal structure, and a layered structure
stacked along the $c$-axis of the hexagonal lattice, consisting of seven atoms in one septuple layer.
The Pb atom is sandwiched by Se-Bi-Se or Te-Sb-Te layers and located at the inversion center, shown
in Fig.1(a). There are van der Waals interactions between two septuple layers, and this provides
a natural surface geometry which is appropriate for observing a topologically protected surface state.

Upon increasing the SOC strength ratio, $\lambda_{\mathrm{SO}}/\lambda_{0}$, from 0 to 1, both materials
show the phase transition from BI to TI; in other words, there are band crossings during the process, cf.,
Fig1.(b)-(g). Critical values of the ratio are 0.46 for PbBi$_2$Se$_4$ and 0.62 for PbSb$_2$Te$_4$,
and the electronic band structures at each critical point are shown in Fig.1(c) and Fig.1(f).
3D Dirac cones are seen at the Z point, one of the time-reversal invariant momentum points where -\textbf{k}
is equivalent to \textbf{k}; they are doubly degenerate due to spatial inversion and time reversal symmetry.

Without the spin-orbit interaction, the system should have a trivial Z$_2$ topological number ($\nu=0$),
the so called BI. The presence of band crossings between conduction and valence bands during increasing
SOC represents the change of its Z$_2$ topological number from trivial ($\nu=0$) to non-trivial ($\nu=1$)
insulators. If the topological phase transition occurs before the SOC strength reaches $\lambda_{0}$,
the system is classified as TI. In these inversion symmetric systems, band crossing is equivalent
to band inversion where conduction and valence bands exchange their parities at the crossing point.

From our calculations, the bulk gaps of PbBi$_2$Se$_4$ and PbSb$_2$Te$_4$ are 0.28eV and 0.13eV, respectively.
The gap size of PbBi$_2$Se$_4$ is comparable to that of Bi$_2$Se$_3$~\cite{Xia2009,Zhang2009}.
The gap values from the calculations without SOC are 0.24eV (Fig.1(b)) and 0.26eV (Fig.1(e)),
which reflects purely hybridization effects. After increasing the SOC strength, gaps shrink to zero and
then increase again, which shows the competition between hybridization and SOC strength in the topological
phase transition. A non-trivial topological phase is a result of the predominance of SOC over the hybridization
strength. Compared to other TI chalcogenides, the large band gap in PbBi$_2$Se$_4$ originates from
weak hybridization and large SOC strength.

To estimate the gap size of PbBi$_2$Se$_4$ more precisely, we performed non-local screened-exchange
LDA (sX-LDA) calculations~\cite{PhysRevB.41.7868,PhysRevB.59.7486} which overcome the well-known weakness
of gap underestimation in the LDA scheme and reproduces the experimental gap values in many semiconducting
materials. Significantly, the sX-LDA result yields a gap of 0.40eV in PbBi$_2$Se$_4$, which confirms
that PbBi$_2$Se$_4$ is one of the largest gap TI, leading to the fact that PbBi$_2$Se$_4$ may be the
best for high temperature applications among the known TI materials.
%In spite of the larger gap in the sX-LDA result, however, it would affect little on the critical SOC value
%because of the small change in the gap compared to the effect of SOC.

%\emph{\textbf{TI to BI transition in Pb$_n$Bi$_2$Se$_{n+3}$ and Pb$_n$Sb$_2$Te$_{n+3}$ series}}

One advantage of the method, used in this work to verify the topological phase, is that it describes
the exact distance from the phase boundary of BI and TI in terms of the SOC strength. For instance,
the larger critical SOC strength of PbSb$_2$Te$_4$ than that of PbBi$_2$Se$_4$ means that the former
is closer to the phase boundary than the latter.

In addition to PbBi$_2$Se$_4$ and PbSb$_2$Te$_4$, we suggest the series of Pb$_n$Bi$_2$Se$_{n+3}$
and Pb$_n$Sb$_2$Te$_{n+3}$ structures where $n$ is an integer larger than 1 and show that a phase
transition from TI to BI occurs as $n$ increases. The crystal structure of Pb$_n$Bi$_2$Se$_{n+3}$/
Pb$_n$Sb$_2$Te$_{n+3}$ is composed of (PbSe)$_n$/(PbTe)$_n$ core with Se-Bi-(Se)/Te-Sb-(Te) sandwich layers.
%As we have already shown, $n=1$ compositions are TI, and $n=\infty$ (PbSe and PbTe) are known as BI.
%Therefore, a topological phase transition should occur between the two ends.
As shown in Fig.2(a), the critical SOC ratio ($\lambda^{\mathrm{c}}_{\mathrm{SO}}/\lambda_{0}$) exceeds 1
for $n\geq$ 3 for both series of materials, which means that there is a phase boundary between $n=2$ (TI)
and $n=3$ (BI). The gap sizes decrease up to $n=2$ and increase again in going beyond the phase boundary.
(cf. Fig.2(b)) On the other hand, the calculated gap values without SOC rise monotonically with $n$; this
shows that as $n$ increases, the hybridization strength exceeds SOC, resulting in the topological phase
transition.

In Fig.2(c) and (d), electronic band structures from the slab geometry of the $n=1$ and $n=3$ composition
of Pb$_n$Bi$_2$Se$_{n+3}$, which consist of 6 septuple and 4 hendecuple layers with 78.4\AA~\ and 85.3\AA
~\ respectively, are shown. Consistent with the critical SOC ratio in Fig.2(a), PbBi$_2$Se$_4$ has
a topologically protected surface state, whereas there is no such state connecting the valence
and conduction bands in the Pb$_3$Bi$_2$Se$_6$ slab. This is further evidence of the change in Z$_2$
topological number from 1 to 0 as $n$ increases. The 2D surface state Dirac cone in the PbBi$_2$Se$_4$
slab is robust under the presence of non-magnetic perturbations. Again, the 0.35eV gap originating from
bulk states is useful for high temperature spintronics applications. In other words, a single isolated
surface state Dirac cone is seen in the energy range from -0.03eV below the Fermi level to 0.32eV above.
In the case of the Pb$_3$Bi$_2$Se$_6$ slab, due to the weak van der Waals type interlayer interactions,
no distinct surface state appears from the bulk BI states.

%\emph{\textbf{3D Dirac cone at the phase boundary : presence of monopole}}

Among the Pb-based chalcogenides series, Pb$_2$Sb$_2$Te$_5$ whose critical SOC ratio is 0.96, is located
quite close to the phase boundary, that its Z$_2$ topological number may be easily tuned by changing
external parameters. The 26meV gap in the fully optimized geometry of Pb$_2$Sb$_2$Te$_5$ (Fig.3(a)) is closed
by making a small change in lattice constants. As shown in Fig.3(b), Pb$_2$Sb$_2$Te$_5$ with a $0.5\%$
reduction of the $ab$-lattice constants is positioned exactly at the borderline of TI and BI, whose electronic
band structure shows a gapless 3D Dirac cone at the $\Gamma$ point, realizing monopoles which generate an
inverse-square type Berry curvature in the 3D Brillioun zone. A reduction of the $ab$-lattice constants makes
the hybridization stronger, resulting in equilibration with the SOC strength.
%Compressive stress on the basal plane makes the hybridization strength stronger, resulting in balancing with SOC strength.
The band structure near the Dirac point shows anisotropic dispersions between in-plane and out-of-plane directions,
described by a massless Dirac Hamiltonian
\begin{equation}
H=v_{\perp}\vec{\sigma}_{\perp}\cdot\vec{k}_{\perp}+v_{\parallel}\vec{\sigma}_{\parallel}\cdot\vec{k}_{\parallel}
\end{equation}
where $\vec{\sigma}$ are Pauli matrices, $\vec{k}_{\perp}$ is a momentum vector along the $c$-axis and
$\vec{k}_{\parallel}$ is on the $ab$-plane. Nearly circular cross-section on the ($k_{x}$-$k_{y}$) plane and
an ellipsoidal cross-sections on the ($k_{z}$-$k_{x}$) plane are depicted in Fig.3(c) and (d), where the
electronic band dispersion and the iso-energy contours of the 3D Dirac cone are shown.

The substitution of Te by Se might be useful to control the topological phase of this material in terms of varying
both the SOC strength and lattice parameters. Considering the smaller ionic radius and SOC strength of Se than
those of Te, the Se substitution can play a role in reducing not only lattice constants but also the SOC strength
which can both accelerate the phase transition from TI to BI. Hence, Pb$_2$Sb$_2$(Te$_{1-x}$Se$_x$)$_5$ might be
appropriate for studying the topological phase transition and the 3D Dirac cone.
%potential candidates?

%\emph{\textbf{Conclusions}}

In this work, we predicted new TI materials of Pb-based chalcogenides by investigating the change of the Z$_2$
topological number during variation of the SOC strength. With increasing $n$ in the Pb$_n$Bi$_2$Se$_{n+3}$ and
Pb$_n$Sb$_2$Te$_{n+3}$ series, the TI to BI transition occurs between $n=2$ and 3. Among these compounds,
Pb$_2$Sb$_2$Te$_5$ has a critical ratio of SOC strength close to 1, which means closeness to the phase boundary.
Both topological phases and the gap at the 3D Dirac cone are tunable by changing lattice constants or the SOC
strength. Also, it will be interesting to investigate the physical properties of the bulk 3D Dirac cone manifested
at the critical point. As an extension of the present work, the Pb$_n$(Bi/Sb)$_{2m}$(Se/Te)$_{n+3m}$ series, which
are pseudo-binary [Pb(Se/Te)]$_n$[(Bi/Sb)$_2$(Se/Te)$_3$]$_m$ systems and combinations of Pb$_n$(Bi/Sb)$_2$(Se/Te)$_{n+3}$
and [(Bi/Sb)$_2$(Se/Te)$_3$]$_m$ layered structures, are potential candidates for new TI materials which might show
various bulk band gaps and topological phase transitions. The method used in our work can be the most appropriate
tool to determine the Z$_2$ topological number of the systems without inversion symmetry, which show pair creation
and annihilation of monopole and anti-monopoles at the topological phase transition region~\cite{PhysRevB.78.165313}.

\begin{acknowledgments}
Support from the U.S. DOE under Grant No.DE-FG02-88ER45372 is gratefully acknowledged.
\end{acknowledgments}

%\cite{*}

%\bibliography{new_candidates}% Produces the bibliography via BibTeX.

\begin{figure}[b]
 \centering\includegraphics[width=17.2cm]{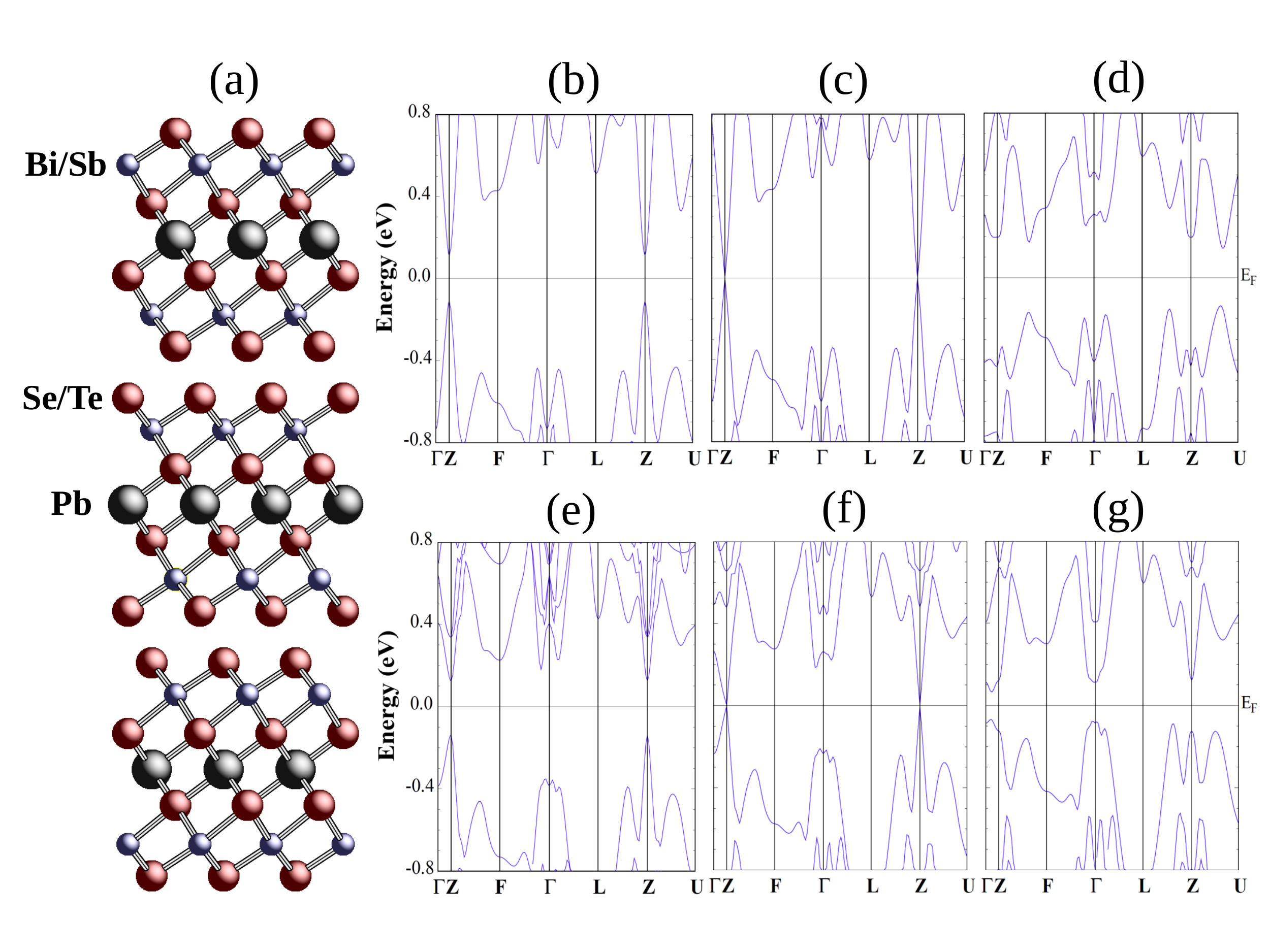}
 \caption{(a) Crystal structures of PbBi$_2$Se$_4$/PbSb$_2$Te$_4$. Electronic band structures of
 (b)-(d) PbBi$_2$Se$_4$ and (e)-(f) PbSb$_2$Te$_4$. Calculations (b),(e) without spin-orbit interactions
 ($\lambda_{\mathrm{SO}}=0$), (d),(g) with real SOC strength ($\lambda_{\mathrm{SO}}=\lambda_0$),
 and (c),(f) with critical SOC strength ($\lambda_{\mathrm{SO}}=\lambda^{\mathrm{c}}_{\mathrm{SO}}$).} \label{fig:1}
\end{figure}

\begin{figure}[b]
 \centering\includegraphics[width=17.2cm]{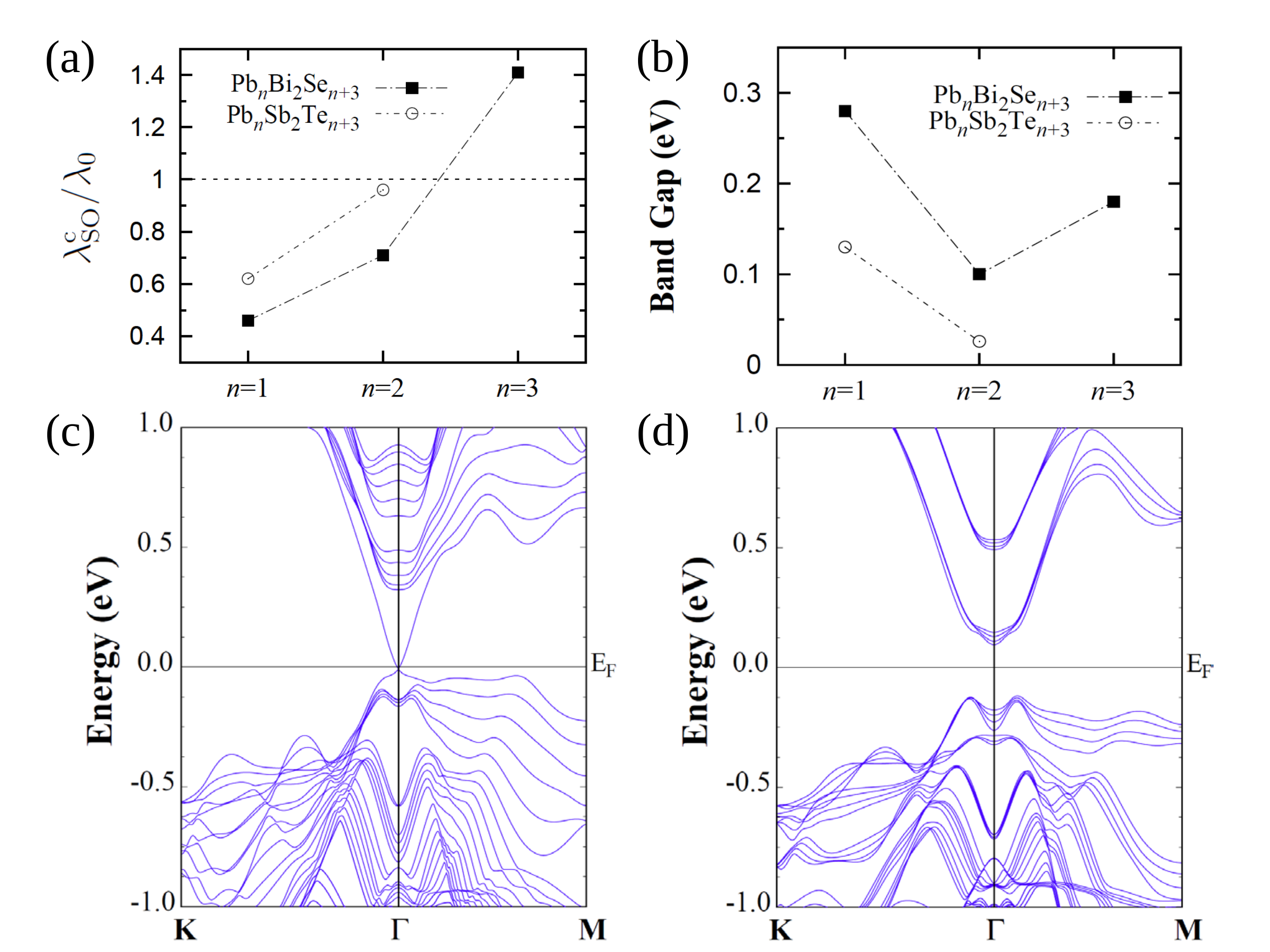}
 \caption{(a) Critical SOC ratio and (b) bulk band gap with respect to $n$ for the Pb$_n$Bi$_2$Se$_{n+3}$
 and Pb$_n$Sb$_2$Te$_{n+3}$ series. Band structures of slab geometry of (c) 6 septuple layers of PbBi$_2$Se$_4$
 and (d) 4 hendecuple layers of Pb$_3$Bi$_2$Se$_6$, which represent TI and BI, respectively.} \label{fig:2}
\end{figure}

\begin{figure}[b]
 \centering\includegraphics[width=17.2cm]{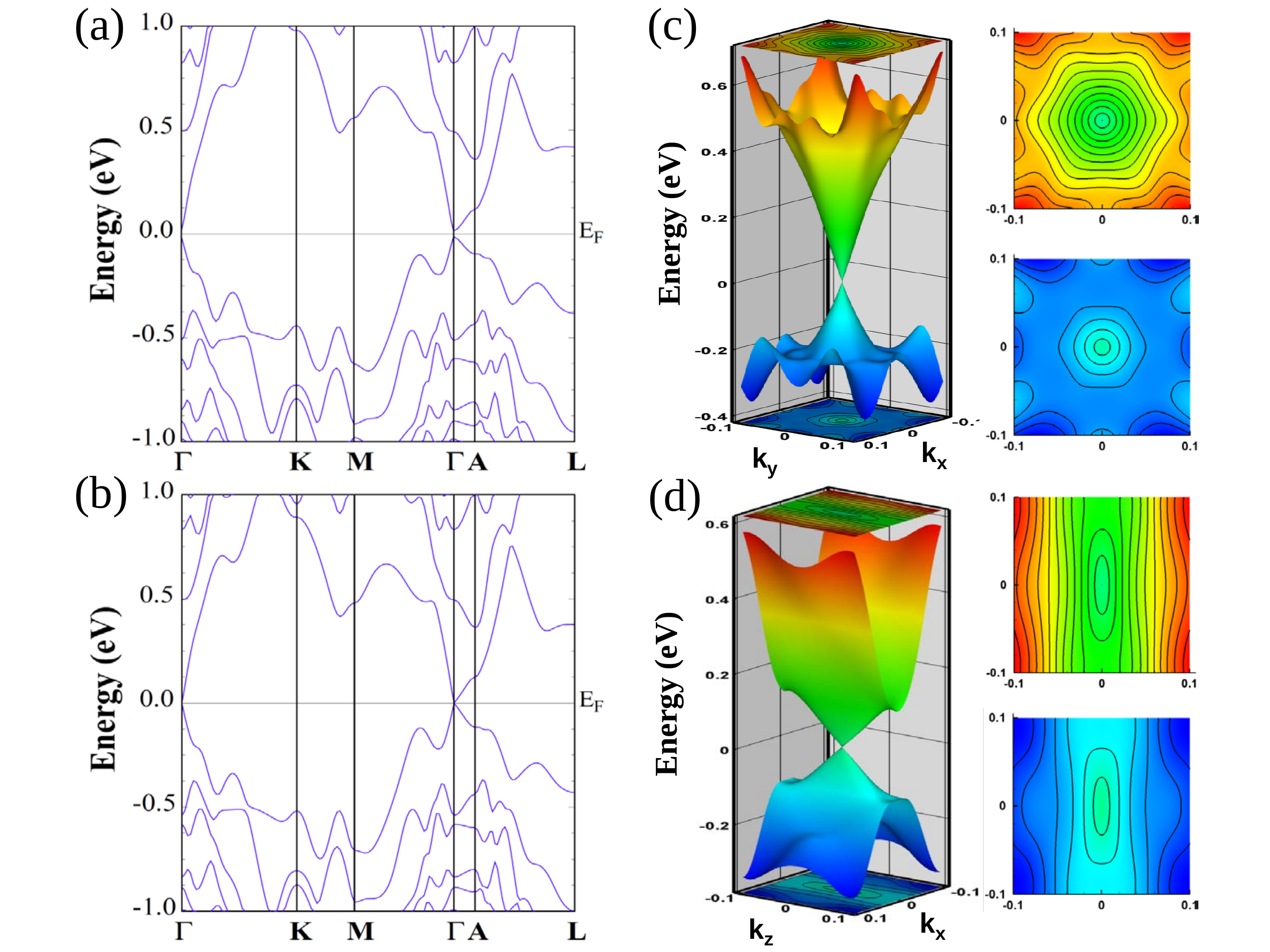}
 \caption{Band structures of Pb$_2$Sb$_2$Te$_5$ with (a) fully optimized crystal structure and (b) 0.5$\%$
 reduction of $ab$-lattice constants. Three-dimensional plot of Dirac cone in (b) and its contours
 are shown in (c) and (d). Anisotropic Dirac cone dispersions on (c) $k_{x}$-$k_{y}$ and (d) $k_{z}$-$k_{x}$
 plane.} \label{fig:3}
\end{figure}

\end{document}